\documentclass[useAMS,usenatbib]{mn2e}
\usepackage{mathrsfs} \usepackage{graphicx}\usepackage{color}
\usepackage{mathptmx} \usepackage{helvet} \usepackage{courier} 
\usepackage{type1cm}  \usepackage{makeidx}  \usepackage{multicol} 
\usepackage[bottom]{footmisc}  \usepackage{bm} 
\usepackage{mediabb}
\def\be{\begin{equation}}\def\ee{\end{equation}}
\def\/{\over} \def\x{\times} \def\p{\propto} 

 \def\deg{^\circ}
  
\def\sin{{\rm sin}\ }  \def\cosec{{\rm cosec}}
\def\ne{n_{\rm e}}\def\xe{x_{\rm e}}
   
\def\la{\langle} \def\ra{\rangle}
\def\({\left(} \def\){\right)} \def\[{\left[} \def\]{\right]} 
 \def\cc{{\rm cm^{-3}}}
\def\Te{T_{\rm e}}  \def\K{{\rm K}}
\def\GHz{{\rm GHz}}  

\def\xe{x_{\rm e}} \def\ne{n_{\rm e}}\def\nHI{n_{\rm HI}} \def\NHI{N_{\rm HI}} 
\def\Ncr{N_{\rm CR}} \def\Tff{T_{\rm ff}} \def\Tsyn{T_{\rm syn}} 
\def\uRM{{\rm rad \ m^{-2}}} \def\uHI{{\rm cm^{-2}}}
\def\uNHI{10^{21}\ {\rm cm^{-2}}} \def\uEM{{\rm pc\ cm^{-6}}}
\def\unHI{{\rm cm^{-3}}} \def\Bpara{B_{//}} \def\muG{\mu{\rm G}}  
\def\Btot{B_{\rm tot}} \def\Bperp{B_{\perp}} 
\def\avB{\la \Bpara \ra}\def\avBpara{\la \Bpara \ra}
\def\avBperp{\la \Bperp \ra} 
\def\avBtot{\la \Btot \ra}\def\avBt{\la \Btot \ra}
\def\avne{\langle \ne \rangle} \def\avnHI{\langle \nHI \rangle} 
\def\avnhi{\langle \nHI \rangle}   \def\pc{{\rm pc}}
\def\Ecr{E_{\rm cr}}
\def\ep{\epsilon}
\def\mH{m_{\rm H}}
\def\HI{H{\small I} }

\def\zhalf{z_{1/2}}
\def\zdisk{z_{1/2:{\rm disk}}}
\def\L{L}
\def\BQ{\mathcal{B_{\rm tot}}}

\title[Magnetic field in the local Galactic disc]{Magnetic field and ISM in the local Galactic disc} 
\author[Y. Sofue et al]{Y. Sofue$^1$\thanks{E-mail:sofue@ioa.s.u-tokyo.ac.jp} 
H.  Nakanishi$^2$, and K. Ichiki$^3$,  \\
1. Institute of Astronomy, The University of Tokyo, Tokyo 181-0015, Japan\\
2. Graduate Schools of Sci. and Engineering, Kagoshima Univ., Kagoshima 890-8544, Japan\\
3. Graduate School of Science, Div. Particle and Astrophys. Sci., Nagoya University,  Nagoya 464-8602, Japan }
\begin{document} 
\date{}   
\maketitle   
\begin{abstract}
Correlation analysis is obtained among Faraday rotation measure, \HI column density, thermal and synchrotron radio brightness using archival all-sky maps of the Galaxy. A method is presented to calculate the magnetic strength and its line-of-sight (LOS) component, volume gas densities, effective LOS depth, effective scale height of the disk from these data in a hybrid way. Applying the method to archival data, all-sky maps of the local magnetic field strength and its parallel component are obtained, which reveal details of local field orientation. 
\end{abstract}

 
\section{Introduction}  

Large scale mappings of Faraday rotation measure (RM) of extragalactic linearly polarized radio sources have been achieved extensively in the decades (Taylor et al 2009; Oppermann et al. 2012), with which various analyses have been obtained to investigate structures of galactic as well as intergalactic magnetic fields (e.g., review by Akahori et al. 2018).

Local magnetic fields in the Solar vicinity have been also studied using these RM data as well as polarization observations of the Galactic radio emission (Mao et al. 2012; {Wolleben et al.} {2010}; Stil et al. 2011; Sun et al. 2015; Sofue and Nakanishi 2017; {Liu et al.} {2017}; {Van Eck et al.} {2017} Alves et al. 2018).  

Synchrotron radio emission is a tool to measure the total strength of magnetic field on the assumption that the magnetic energy-density (pressure) is in equipartition with the thermal and cosmic ray energy densities (e.g., Sofue et al. 1986). This method requires information about the depth of emitting region in order to calculate the synchrotron emissivity per volume, as the intensity is an integration of the emissivity along the line of sight (LOS).

Rotation measure is an integration of the parallel component of magnetic field multiplied by thermal electron density along the line-of-sight (LOS). It is related not only to thermal (free-free) radio emission, but also to \HI column density through thermal electron fraction in the neutral interstellar medium (ISM).

Determination of the LOS depth is, therefore, a key to measure the magnetic strength from synchrotron emission and the parallel magnetic component from RM. The depth is also required to estimate the volume densities of \HI and thermal electrons from observed \HI and thermal radio intensities. Emission measure and \HI column density are useful to estimate the LOS depth, given a relation between the thermal and \HI gas densities is appropriately settled.

In this paper, correlation analyses are obtained among various radio astronomical observables (RM, \HI column density, thermal and synchrotron radio brightness) in order to determine physical quantities of the ISM such as the magnetic strength, gas densities, and  LOS depth. One of the major goal of the present hybrid analysis method will be to obtain whole-sky maps of the total strength and parallel component of the magnetic field in the local Galactic disk.

\begin{figure} 
\begin{center}      
\includegraphics[width=70mm]{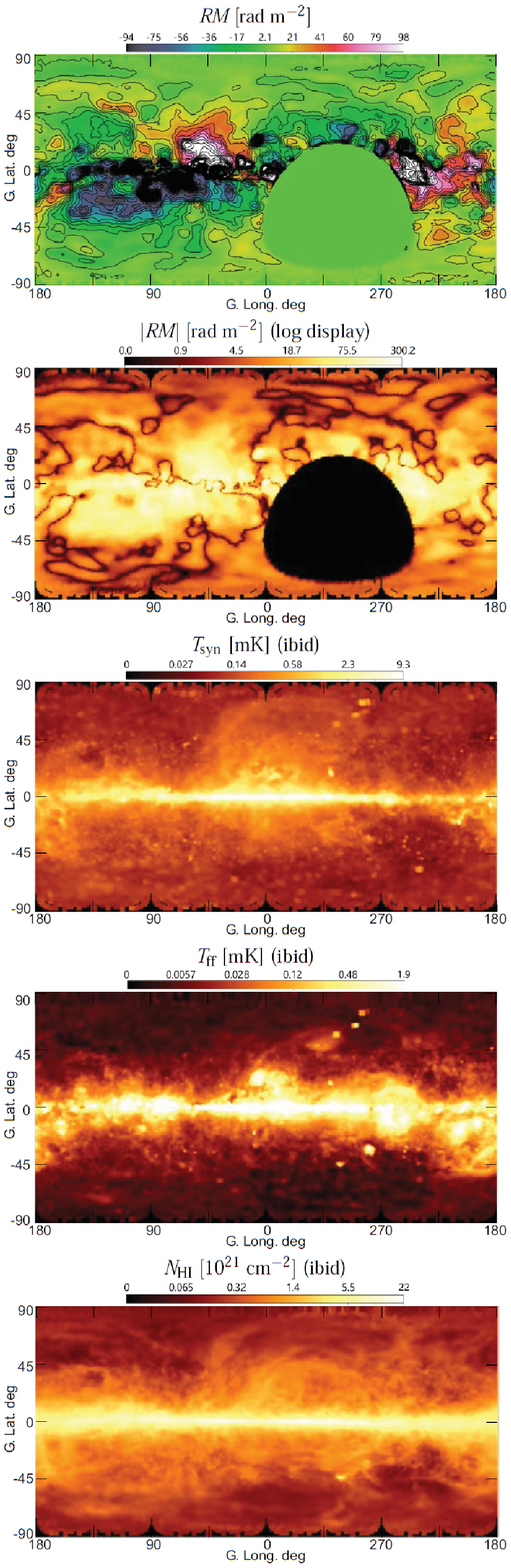}  
\end{center}
\caption{All-sky maps of $RM$, $|RM|$ ($\uRM$)  (Taylor et al. 2009), $\Tsyn$ and $\Tff$ (mK) at 23 GHz (Gold et al 2011), and  $\NHI$ ($\uNHI$) (Kalberla et al. 2005). }
\label{mapData}
\end{figure}

\section{Observables and correlations}

\subsection{Data}

The observational data for the Faraday rotation were taken from all-sky RM survey by Taylor et al. (2009), \HI data from the Leiden-Argentine-Bonn (LAB) survey by Kalberla et al. (2005), synchrotron and free-free emissions at 23 GHz from the 7-years result of Wilkinson Microwave Anisotropy Probe (WMAP) project by Gold et al (2011).  Figure \ref{mapData} shows the employed map data for $RM$,  $|RM|$, \HI column density $\NHI$, thermal (free-free) radio brightness temperature $\Tff$ and synchrotron radio brightness $\Tsyn$, both at 23 GHz. 

\begin{figure*} 
\begin{center} 
\includegraphics[width=13cm]{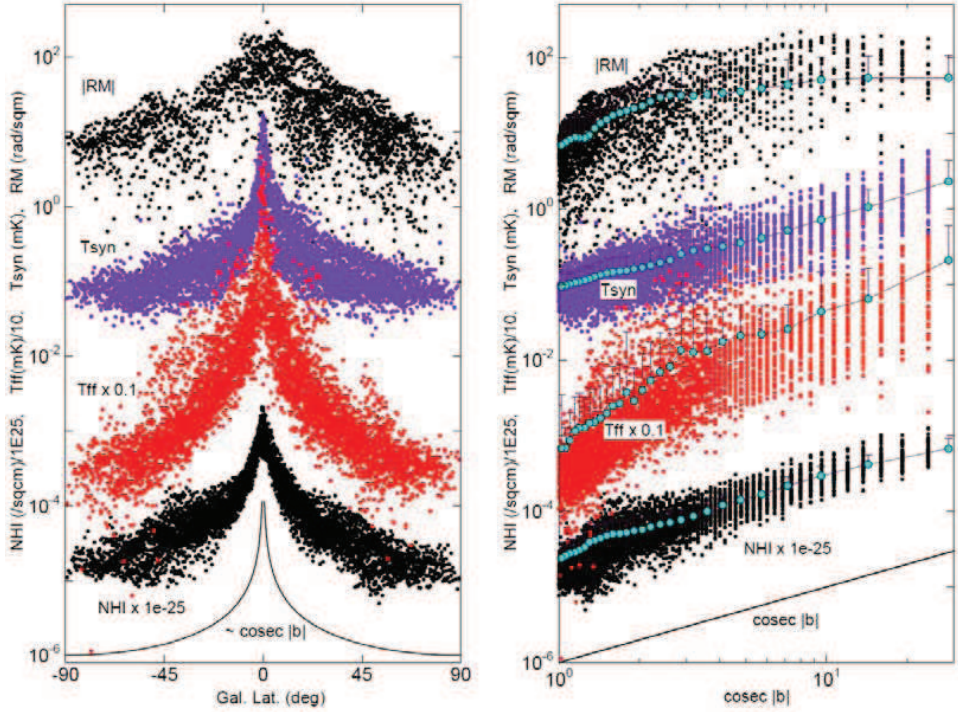}       
\end{center}
\caption{$|RM| \ (\uRM)$, $\Tsyn$ (mK), $0.1\x \Tff$ (mK) and $10^{-25}\x \NHI (\uHI)$  plotted against latitude $b$. Points are reduced by 1 per 10. Note the beautiful dependence of the plotted quantities on cosec $|b|$ relation shown by the full lines, indicating that they are tightly coupled with the line-of-sight depth of a plane parallel layer. 
 Right panel shows the same, but against cosec $|b|$. Open circles are averages of the plots in every $2\deg$ bin of absolute latitude. Bars show standard deviations (sd) of the original values, which are given only in upper sides in order to avoid logarithm of negative values for large sd.
}
\label{All-Lat}
\end{figure*}

Figure \ref{All-Lat} shows plots of the same data in figure \ref{mapData} against the latitude $b$ and cosec $|b|$. The global similarity of the latitudinal variations indicates that the four observed quantities are deeply coupled with each other. It is remarkable that all the plots beautifully obey the cosec $|b|$ relation shown by the full lines. This fact indicates that these radio observed quantities are tightly coupled with the line-of-sight depth (LOS) through galactic disk composed of a plane parallel layer in the first approximation.

A more detailed inspection of the figures reveals that, besides the global common cosec $|b|$ property, there exist systematic differences in the latitudinal variations among the quantities. The rotation measure, $RM$, shows milder increase toward the galactic plane than the other quantities. The saturation of $|RM|$ near the galactic plane suggests that the magnetic field directions are reversing there. On the other hand, synchrotron intensity has much sharper peak at the plane, and shows similar variation to \HI intensity. Another remarkable property is the sharper increase of the thermal emission toward the galactic plane than \HI. This manifests stronger dependence of the thermal emission on the ISM density through the emission measure $EM \p \ne^2 L$ than that of the \HI column $\NHI \p \nHI L$.
 
The radio observables are related to the ISM quantities as follows. Faraday rotation measure $RM$ is related to the thermal electron density $\ne$ and line-of-sight (LOS) component of the magnetic field strength $\Bpara$ through
\be
\(RM\/\uRM \)\sim 0.81\(\la\ne\ra\/\cc\)\(\la \Bpara \ra\/\muG\) \( L\/\pc\). 
\label{RM}
\ee 
The emission measure $EM$ is rewritten by the volume density $\ne$ of thermal electrons as
\be
\(EM\/\uEM \) \sim \( \avne \/ \cc \)^2 \(L \/ \pc \).
\ee
The \textsc{HI} column density $\NHI$ is given by the \HI volume density $\nHI$ as
\be
\(\NHI\/{\rm cm^{-2}}\) 
\sim 3.086\x 10^{18} \(\avnHI\/\unHI\) \(L\/\pc\).
\ee
Here, $\la ~ \ra$ denotes LOS average, and is defined and described later. 
The synchrotron radio brightness $\Sigma_\nu$, as observed by the brightness temperature $\Tsyn$, is related to the volume emissivity $\ep$, frequency $\nu$, and $L$ as
\be
\Sigma_\nu={2k\Tsyn\/\lambda^2} \sim {L\/4\pi}{ d\ep\/d\nu}\sim {L\/4 \pi}{ \ep \/ \nu},
\label{Sigma}
\ee
which may be rewritten in a practical way as
\be
\(\Tsyn \/{\rm K}\) \sim 1.472\times 10^{11}\(\nu \/{\rm GHz}\)^{-3}
\(\ep\/{\rm erg\ cm^{-3}s^{-1}}\)  \(\L\/{\rm pc}\),
\label{Tsynbyep}
\ee 
where $\lambda=c/\nu$ is the wavelength and $k$ is the Boltzmann constant.  

We assume that the Galactic disk is composed of four horizontal layers (disks) of \textsc{HI} gas, thermal electrons, magnetic fields and cosmic rays, which have the same half thickness (scale height) $\zdisk$. This means that the LOS depth $L$ of the four quantities are equal. This assumption may not be good enough for the synchrotron emission that may originate from a thicker magnetic halo. However, it may be considered that the contribution of magnetic halo to $RM$ and synchrotron emission is much smaller than that of the disk because of weaker magnetic strength and electron density by an order of magnitude. The here used depth $L$ is an effective depth, and is related to the geometrical depth through a volume filling factor, as will be described later in detail.

Based on these considerations, we assume that the ISM quantities are smooth functions of the effective LOS depth $L$ for the first approximation, and an average of any quantity $f$ over $L$ satisfies the following relation,
\be
\la f \ra ={\int_0^L f dx\/\int_0^L dx}={\int_0^L f dx \/L}.
\label{f}
\ee
We also assume for any quantities $f$ and $g$
\be
\la f \ra \sim  \la f^2 \ra ^{1/2},
\label{ff}
\ee
and
\be
\la fg \ra \sim \la f \ra \la g \ra.
\label{fg}
\ee  

\subsection{Correlation among Radio Observables}  

\subsubsection{Free-Free to \textsc{HI} tight relation}

\begin{figure*} 
\begin{center} 
\includegraphics[width=16cm]{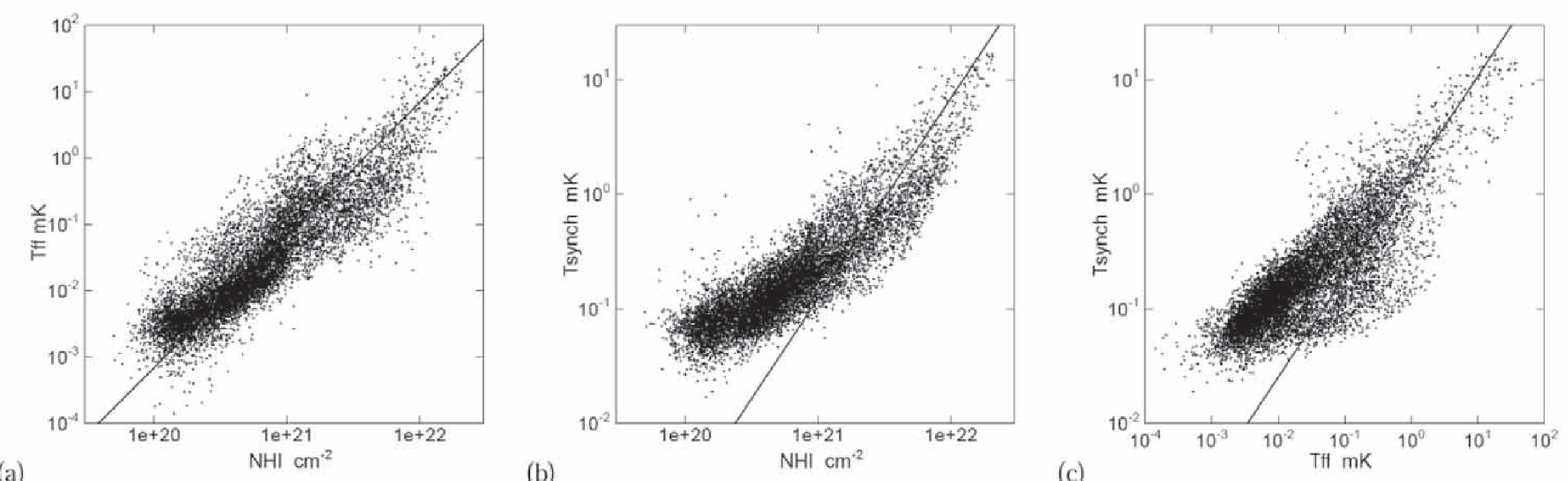}      
\end{center} 
\caption{Correlation of $\Tff$ and $\NHI$ with a line of power index 2, $\Tsyn$ and $\NHI$ with a line of index 8/7.  Vertical scaling of the lines are arbitrary.  
}
\label{ISM-ISM}

\begin{center} 
\includegraphics[width=16cm]{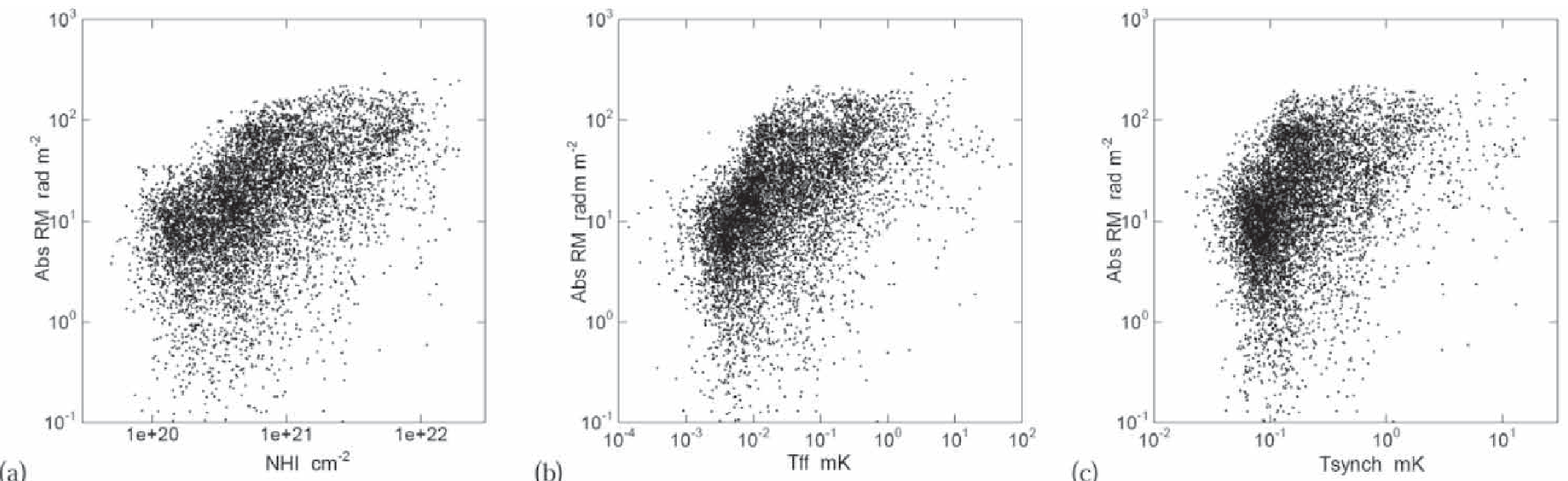}    
\end{center}
\caption{Correlation of $|RM|$ with (a) HI, (b) FF and (c) Synchrotron emissions. }
\label{RM-ISM}
\end{figure*}

Figure \ref{ISM-ISM}(a) shows a plot of $\Tff$ against the square of $\NHI$. The straight line indicates a $\Tff \p \NHI^2$, and plots on the log-log space well obeys this proportionality. This relation indicates that the thermal electron density $\avne$ is approximately proportional to $\avnhi$, if $L$ is not strongly variable from point to point, which is indeed the case except for the high $\NHI$ region close to the galactic plane. This correlation will be used to estimate the electron density from \textsc{HI} column.

\subsubsection{Synchrotron to ISM relation}

Figure \ref{ISM-ISM}(b) is a plot of $\Tsyn$ against $\NHI$. High intensity region is approximately represented  by a power law of index 7/4, $\Tsyn \propto \NHI^{7/4}$, as expected from frozen-in magnetic field into the ISM and energy-density equipartition between the magnetic field, cosmic rays, and ISM  (see Appendix). Plot of $\Tsyn$ against $\Tff$ in (c) shows a similar relation, where $\Tsyn \propto \Tff^{7/8}$ is expected from the equipartition, because $\NHI \propto \Tff^{1/2}$. In both plots, the synchrotron emission tends to exceed the energy equipartition lines at low intensity regions (high latitudes). This yields larger uncertainty of the estimated magnetic strength at high latitudes.

\subsubsection{RM to ISM relation}

Figure \ref{RM-ISM} shows plots of the absolute RM values against (a) $\NHI$, (b) $\Tsyn$ and (c) $\Tff$. It is impressible that the plots are more scattered than those in figure \ref{ISM-ISM}. This is because the rotation measure is an integrated function of the magnetic field strength along the LOS including the reversal of field direction. This scattered characteristics of RM is useful to derive the spatial variation of the LOS field direction and strength $\Bpara$.  

\begin{figure} 
\begin{center} 
\includegraphics[width=7cm]{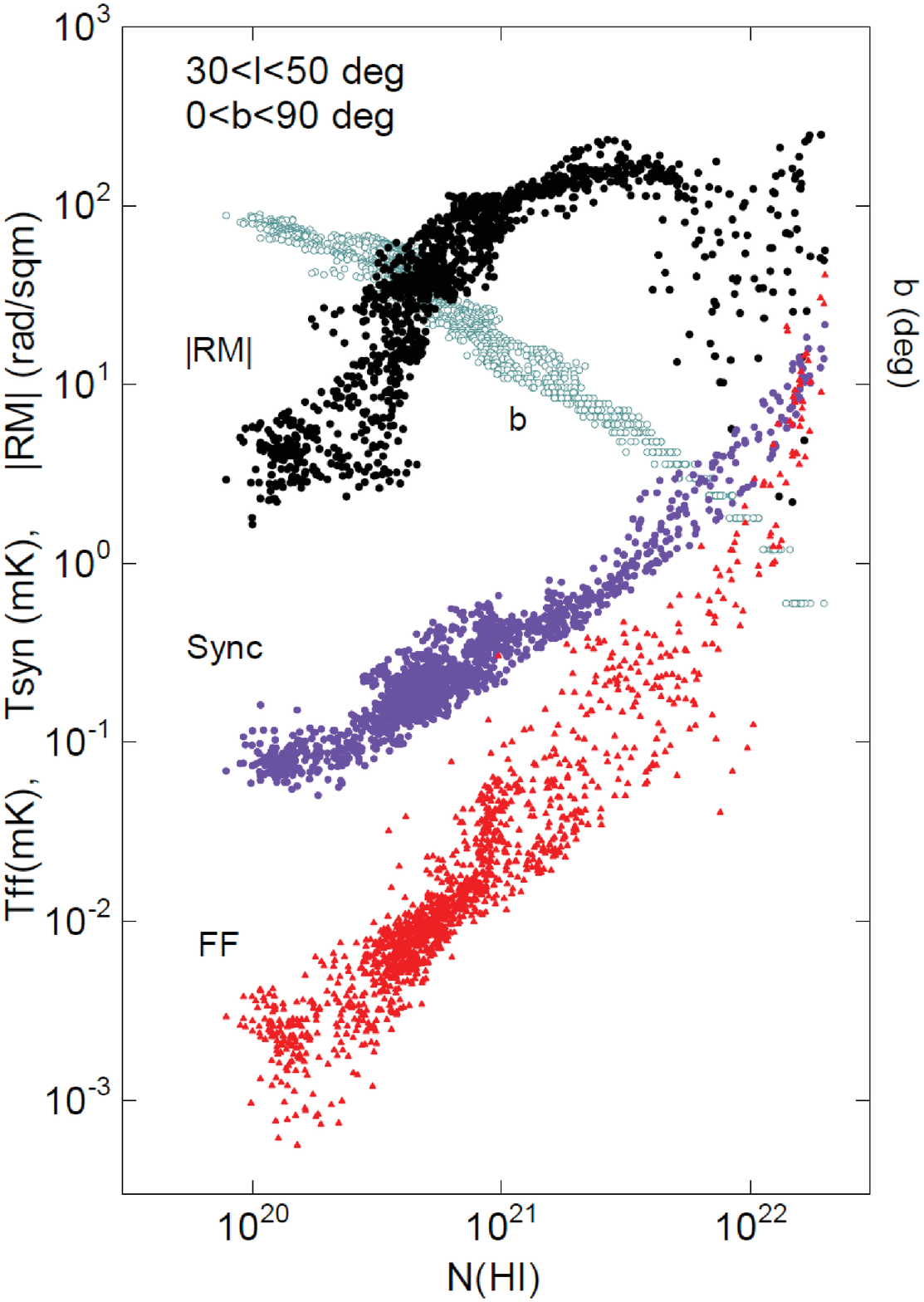} 
\end{center}
\caption{Correlations of $|RM|$, $\Tff$ and $\Tsyn$ with $\NHI$ in a restricted area on the sky ($30\deg\le l \le 50\deg$, $b\ge 0\deg$). Shown by green circles are latitudes  with the values indicated by the right axis}. 
\label{All_to_HI}
\end{figure}

Although $|RM|$ is scattered against the other ISM observables in the whole-sky data, it may better be correlated in a narrower restricted region. Figure \ref{All_to_HI} shows an example of plots of $|RM|$, $\Tsyn$ and $\Tff$ against $\NHI$ in a small area in the 1st quadrant of the Galaxy at $30\deg\le l \le 50\deg$ and $b\ge 0\deg$. Shown by green circles are latitudes corresponding to individual $\NHI$ data points, indicating the tight dependence of $\NHI$ on the latitude through LOS depths. This figure demonstrates how $|RM|$ is tightly correlated to $\NHI$ in a restricted area. 
  
\begin{figure} 
\begin{center}  
\includegraphics[width=7cm]{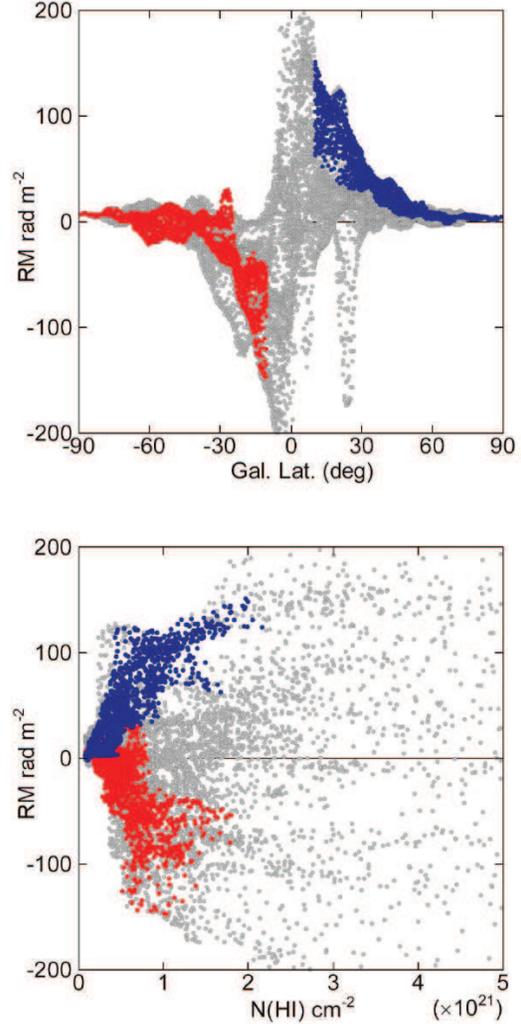}   
\end{center}
\caption{RM against galactic latitude (top) and HI column density (bottom) in the 1st quadrant of the Galaxy. Grey circles show all data points, blue dots are for a small region at $30\deg\le l \le 60\deg$ and $b \ge 10\deg$ and red for $30\deg \le l \le 60\deg$ and $b \le -10\deg$  in the same quadrant.  }
\label{RM-HI-Lat}
\end{figure}

Figure \ref{RM-HI-Lat} shows plots of $RM$ against galactic latitude and \textsc{HI} column in the 1st quadrant of the Galaxy. Grey dots show all points, and blue and red dots represent those in northern and southern two small regions in the same quadrant at $30\deg\le l \le 60\deg$ and $b \ge 10\deg$ and at $30\deg\le l \le 60\deg$ and $b \le -10\deg$. 

Absolute RM value increases toward the galactic plane, the sign of RM changes from negative to positive as the latitude increases, which indicates sudden reversal of the magnetic field direction. The bottom panel in the figure represents the same phenomenon in terms of the column density of \textsc{HI} gas. 

Linear relation of RM with \textsc{HI} column is found in low $|RM|$ and $\NHI$ regions. However, the linearity is lost toward the galactic plane at $|b|\le \sim 10\deg$ with increasing $\NHI$. This represents decrease in the LOS component of magnetic strength $|\Bpara|$ toward the plane, which indicates rapid change of the field direction near the plane.
 
\section{Gas densities, Line-of-sight depth, and disk thickness}

\begin{figure} 
\begin{center}  
\includegraphics[width=5.5cm]{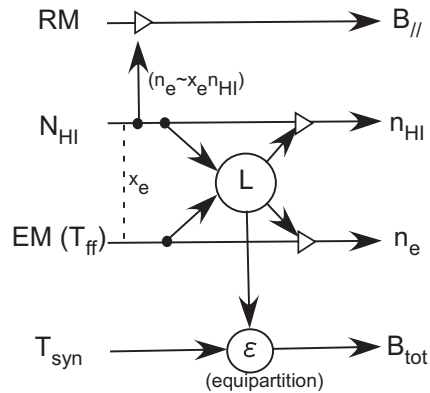} 
\end{center}
\caption{ISM hybrid for determination of physical quantities in the local disk from multiple  radio observables.  }
\label{ISMhybrid}
\end{figure}

Given the four radio observables ($RM$, $\Tff$, $\Tsyn$, and $\NHI$), four ISM parameters (averaged electron density $\avne\sim \xe \nHI$,  effective LOS depth $L$, total magnetic intensity $B$, and LOS component of magnetic field $\Bpara$) can be estimated as follows. Figure \ref{ISMhybrid} illustrates the flow of the analysis, which we call the ISM hybrid.   The effective depth $L$ will be related to geometrical scale height of the disk through volume filling factor.

\subsection{Thermal electron density}

Thermal electron density is assumed to be proportional to the \textsc{HI} gas density as
\be
\ne = \xe \nHI,
\ee
where, $\xe$ is the thermal electron fraction in the neutral ISM,
\be
\xe={\avne\/\avne+\avnhi}\sim {\avne \/\avnhi},
\ee    
which is assumed to be $\xe\sim 0.1$ after Foster et al. (2013), who obtained $\xe\sim 0.08$. 
 Local \textsc{HI} gas is considered to be in cold phase from recent measurement of spin temperature (Sofue 2017, 2018). Even if warm \textsc{HI} is contaminated, its density is an order of magnitude lower, so that the column density is not much affected by warm HI, unless the scale height of warm \textsc{HI} is an order of magnitude greater than that of cold HI.

Electron density is obtained by dividing the emission measure by column density of electrons, which is related to \textsc{HI} column, or $\ne \sim (\ne^2 L)/(\ne L) \sim (\ne^2L)/(\xe \nHI L)\sim EM/(\xe \NHI)$. Thus, we have  
\begin{eqnarray}
\({\avne\/\cc}\) 
\sim 3.09\x 10^{-3}\({EM\/\uEM}\) \(\xe\NHI\/\uNHI\)^{-1}\\ \nonumber
\sim 6.82\times 10^2 \xe^{-1} \(\NHI\/\uNHI\)^{-1} \(\Tff\/{\rm K}\)_{23{\rm GHz}}
\label{avne}
\end{eqnarray}
at $\nu=23$ GHz, where the following relations were used.
The emission measure is expressed by $\Tff$, electron temperature ($\Te \sim 10^4$ K), and observing frequency ($\nu=23$ GHz) as
\begin{eqnarray} 
\(EM \/\rm pc \ cm^{-6}\)  
= {3.05\x 10^2} \(\Tff\/K\) \( \Te \/10^4\K \)^{0.35}\( \nu \/\GHz \)^{2.1} \\ \nonumber 
 \sim 2.21\x 10^5\({\Tff\/\K}\)_{23 \GHz}, ~~~~~~~~~~~~~~~~~~~~~~~~~~~~~~~~  
\label{EMinTb}
\end{eqnarray}
for $\Te=10^4$ K and $\nu=23$ GHz, where optical depth is given by (Oster 1961),
\be
\tau=3.28\x 10^{-7} \( \Te \/10^4\K \)^{-1.35}\( \nu \/\GHz \)^{-2.1}\(EM \/{\rm pc \ cm^{-6}}\),
\ee
which is related to $\Tff$ for optically thin case as
\be
\Tff =(1-e^{-\tau})\Te \simeq \tau \Te.
\ee 

\begin{figure} 
\begin{center}  
\includegraphics[width=85mm]{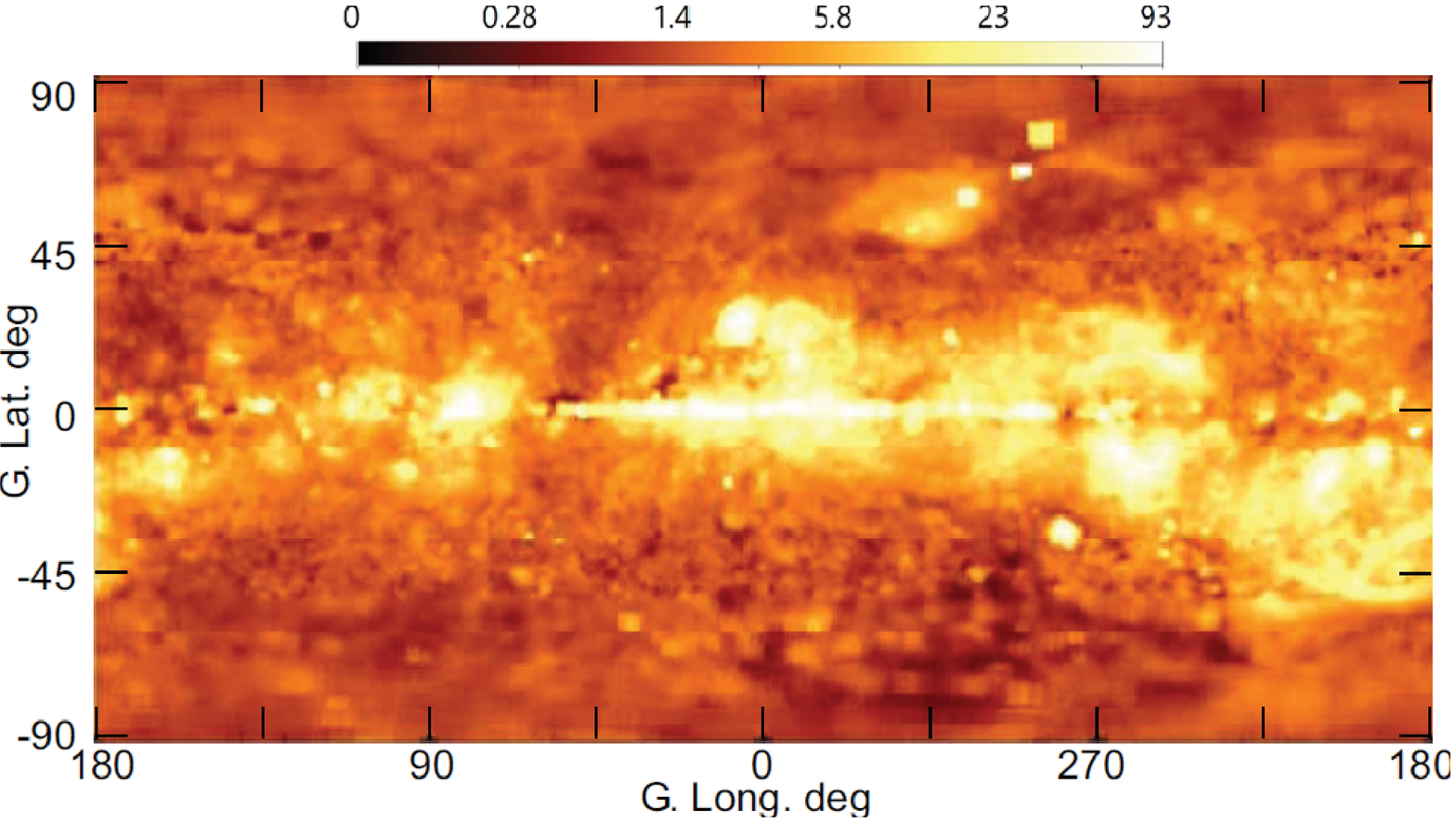}     
\end{center}
\caption{Logarithmic display of the volume density map of \textsc{HI} and thermal electrons. Shown is $\avnHI\ [\cc]$, which is assumed to be proportional to thermal electron density as $\avne \sim \xe \avnHI\ [\cc]$ with $\xe \sim 0.1$.  }
\label{mapDen}
\end{figure}

Figure \ref{mapDen}  shows a calculated all-sky map of $\nHI$, which is equal to $\ne/\xe$, using equation (\ref{avne}). The derived \textsc{HI} density at $|b|>\sim 10\deg$ has nearly a constant value around $\sim 1-2\ \cc$, except for clumpy regions and the GC.  
 
\subsection{LOS depth, scale height, and volume filling factor}

\begin{figure} 
\begin{center} 
 \includegraphics[width=7cm]{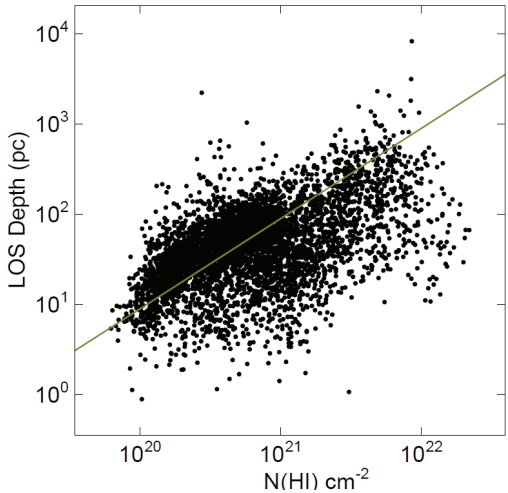}   
\end{center}
\caption{Effective LOS depth $L$ of HI gas for $\xe=0.1$ plotted against $\NHI$. The straight line indicates a linear relation, $\NHI \propto L$, with arbitrary vertical scaling.  }
\label{L-NHI}
\end{figure}

Recalling that $\NHI\sim L \nHI$ and $EM\sim L\ne^2 \sim L\xe^2\nHI^2$, the LOS depth $L$ is given by 
\begin{eqnarray}
\({L\/\pc}\) 
\sim 1.05 \x 10^5 \( \xe \NHI \/ \uNHI \)^2\(EM\/\uEM\)^{-1} \nonumber \\
\sim  0.475\({\xe\NHI \/ \uNHI}\)^2\({\Tff\/\K} \)^{-1}_{23 \GHz}.
\label{L_in_Tff}
\label{L}
\end{eqnarray}
Figure \ref{L-NHI} shows the derived $L$ plotted against $\NHI$ for $\xe=0.1$. The plot roughly obeys the linear relation, $\NHI \propto L$, indicated by the straight line, while points are largely scattered.

\begin{figure} 
\begin{center}  
\includegraphics[width=7cm]{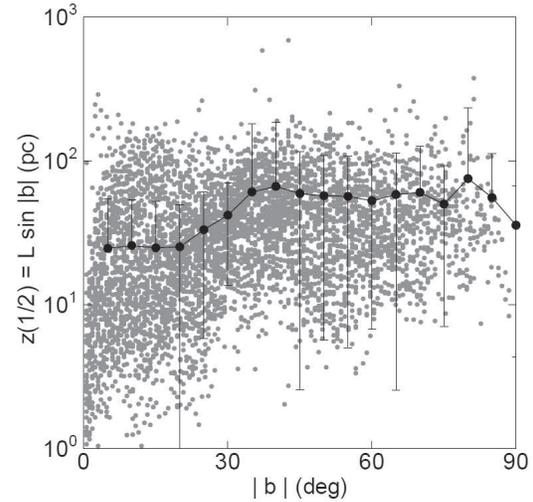}     
\end{center}
\caption{"Effective" half thickness (scale height) of the \textsc{HI} disk defined by $\zhalf=L\ \sin \ |b|$ plotted against latitude, which is related to geometrical half thickness $\zdisk$ by $\zhalf \sim \eta^{1/3} \zdisk$ with $\eta$ being the volume filling factor of th gas. Circles are averages of the neighboring latitude points with standard errors by bars. Points without bar indicate errors greater than the averaged values.  }
\label{z-b}
\end{figure}
 
The here defined $L$ is an "effective (physical)" LOS depth, and is related to the effective half thickness (scale height) of the disk, $\zhalf$, by
\be
L=\zhalf \cosec \ |b|.
\label{L}
\label{Ltozdisk}
\ee
The effective half thickness is further related to the 'geometrical' half thickness
\be
\zhalf=\eta^{1/3}\zdisk,
\label{L}
\label{Ltozdisk}
\ee
where $\eta$ is the volume filling factor of the ISM. The factor $\eta$ will be determined using these relations referring to independent measurement of the \textsc{HI} disk scale height.  

Figure \ref{z-b} shows calculated $\zhalf$ against latitude. Original data points are shown by gray dots, and averaged values in latitudinal interval of $\Delta b= \pm 5\deg$ are shown by big dots with standard errors. Points without error bar indicate those, whose errors are greater than the averaged values. The averaged effective half thickness tends to a constant of $\zhalf\sim 60\pm 8$ pc at $40\deg \le |b| \le 80\deg$.

From the current measurements of \HI disk half thickness 
(160 pc, Lockman  (1984); 
150 pc, Wouterloot et al. (1990);
200 pc, Levine et al. (2006); 
173 pc, Kalberla et al. (2007); 
200 pc, Nakanishi and Sofue (2016); 
217 pc, Marasco et al. (2017)),  
we adopt a simple average of the authors' values, $\zdisk=183\pm 26$ pc. In order for the present determination of $\zhalf$ to satisfy equation (\ref{L}), we obtain $\eta \sim (\zhalf/\zdisk)^3=0.035\pm 0.007$. This value agrees with the recent determination for cold HI gas by Fukui et al. (2018). 
 
\section{Magnetic Fields}

\subsection{Parallel component}

The LOS (or parallel) component of the magnetic field can be obtained by dividing RM by the column density of thermal electrons as
\be
\({\avB \/\muG}\) 
\sim 3.81\x 10^{-3}\({RM\/{\rm rad\ m^{-2}}}\) \(\xe\NHI\/\uNHI\)^{-1}.
\label{Bpara}
\ee 
It is stressed that this formula yields $\avB$ directly from the observables $RM$ and $\NHI$ without employing the LOS depth $L$. So, $\avB$ is the most accurate quantity determined in this paper. 

Equation (\ref{Bpara}) is particularly simple and useful to estimate the parallel component of magnetic strength, because it includes only two observables, $RM$ and $\NHI$, where the effective LOS depth $L$ has been canceled out, leaving $\xe$ as one parameter to be assumed. This relation is now applied for mapping of the $\Bpara$ value on the sky assuming $\xe=0.1$.

First, the sky is binned into $1\deg$ degree grids in longitude and latitude, and calculate averages of RM and $\NHI$ values within $\pm 1\deg$ about each grid point for $|b|\le 50\deg$ region, and within $\pm 2\deg$ for $|b|>50\deg$. At each point on the grids, $\avB$ is calculated with the aid of equation (\ref{Bpara}). By this procedure a $361\x 181$ meshed map of $\avB$ is obtained on the sky with a resolution of $1-2\deg$.

\begin{figure*}  
\begin{center}
\includegraphics[width=13cm]{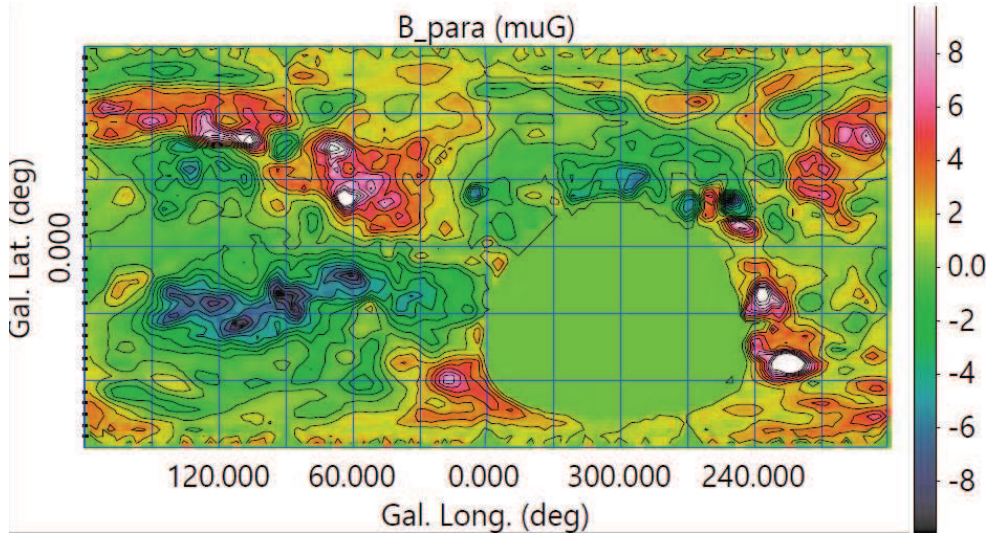}     
\end{center}
\caption{All-sky maps of  $\avBpara$ with contours at interval of $1$ $\muG$.  Positive value with red color indicates a field away from the observer, and negative with blue, approaching. }
\label{mapBpara}
\end{figure*}

Figure \ref{mapBpara} shows the thus obtained all-sky map of $\avB$.
  Positive value in red color indicates a magnetic field away from the observer, and negative with blue indicates a field approaching the observer. 

Let us remember that the RM map was strongly affected by the peaked line-of-sight depth near the galactic plane, causing large positive and negative values near the plane. This caused steep latitudinal gradient of RM due to the field reversal from north to south, resulting in RM singularity along the galactic plane.

On the other hand, the $\avB$ map is not affected by the LOS depth, so that it exhibits the field strength and direction only, so that the RM singularity along the galactic plane does not appear. The map reveals a widely extended arched region with positive magnetic strength of $\avB\sim +5 \muG$ in the north from $(l,b)\sim (40\deg,5\deg)$ to $(210\deg,0\deg)$. This arch seems to be continued by a negative strength arch with $\avB\sim -5\muG$ in the south from $(l,b)\sim (50\deg,-5\deg)$ to $(160\deg,-30\deg)$. 

It may be possible to connect the positive and negative $\Bpara$ arches to draw a giant loop, or a shell, from $l\sim 40\deg$ to $220\deg$ with the field direction being reversed from north to south. Alternatively, the positive arch may be traced through the empty sky around the south pole in the present data (Taylor et al. 2009), where the improved map shows positive RM (Oppermann et al. 2012). If this is the case, the RM arches may trace a sinusoidal belt from the southern hemisphere in the 1st and 2nd quadrants to northern in the 3rd and 4th quadrants, drawing an $\infty$ shaped belt on the sky, with the necks in the galactic plane at $l\sim 30\deg$ and $240\deg$. 
The arched magnetic region along the Aquila Rift from $(l,b)\sim (30\deg,-10\deg)$ to $(300\deg,+30\deg)$ with $\avB\sim +2$ to $-4 \muG$ could be a part of the $\infty$ belt.

 It is also interesting to note that both the northern and southern polar regions show positive $\Bpara$ with $\avB\sim +1 \ \muG$, indicating that the vertical (zenith) field directions are pointing away from the Sun.
 
\begin{figure*}  
\begin{center}  
\includegraphics[width=13cm]{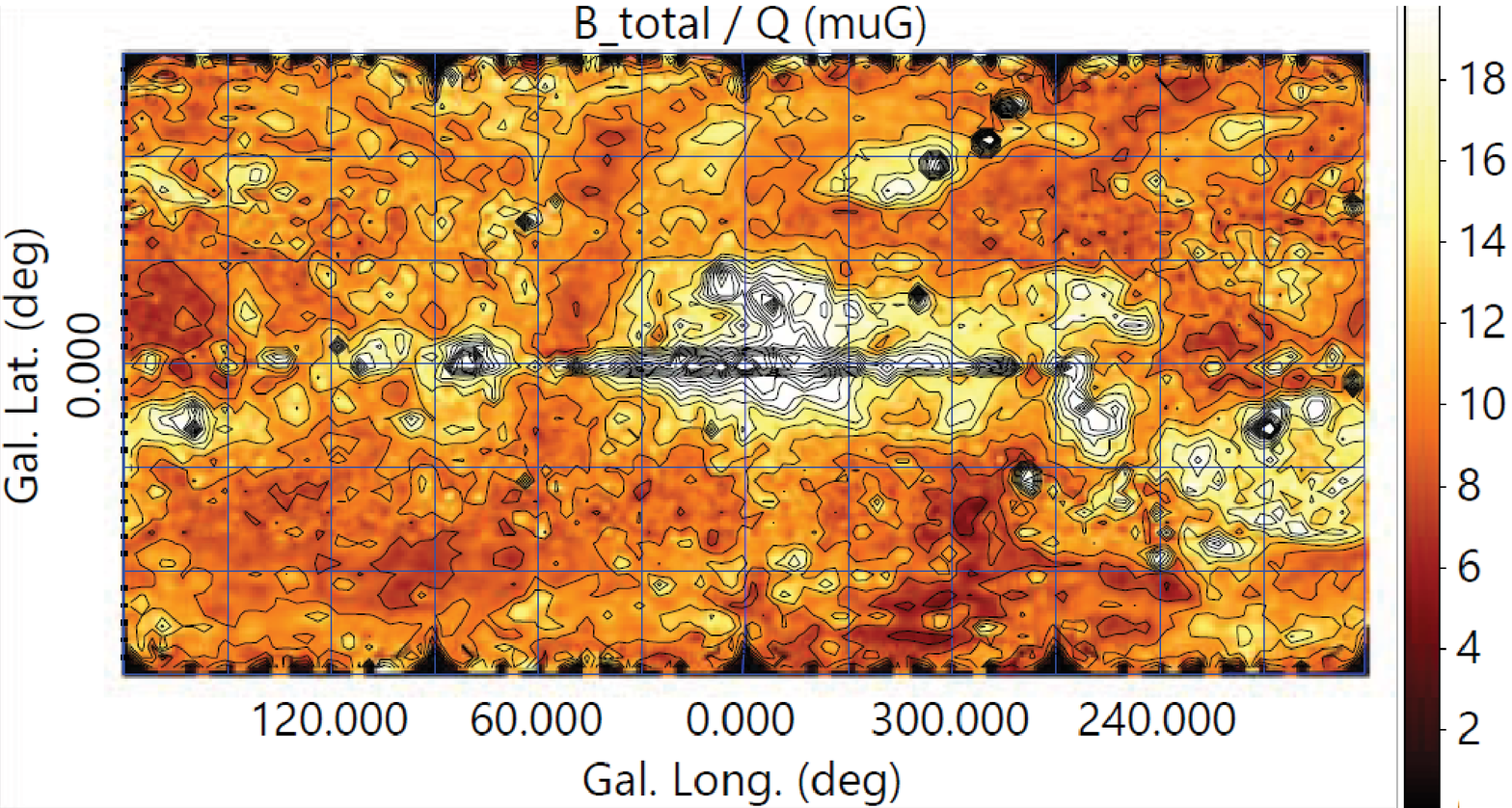}   
\end{center}
\caption{All-sky maps of  $\avBtot/Q$ with contours at interval of 2 $\muG$. The $\eta$-corrected strength is obtained by $\BQ\sim 0.72\times$ map values.}
\label{mapBtot}
\end{figure*}

\subsection{Total  intensity}

The total magnetic intensity $\Btot$ is calculated by assuming that the magnetic and cosmic ray energy densities are in equipartition as (see Appendix)
\be
B^2/8\pi \sim \Ncr \Ecr,
\label{Bequip} 
\ee 
where $\Ncr$ is the cosmic-ray electron number density and $\Ecr$ is representative energy of radio emitting cosmic rays. The magnetic strength is then related to the frequency $\nu$ and volume emissivity $\ep$ as
\be
\(\avBt\/ \muG \)
\sim 3.03 \x 10^8   Q \({\nu\/\GHz}\)^{-1/7}{\ep}^{2/7},
\label{B}
\ee  
where, $Q$ is an equipartition factor, which depends on various assumed conditions and source models. There have been decades of discussion about $Q$ since Burbidge (1956), which includes dependence on such parameters as the spectral index, cut-off frequencies, proton-to-electron density ratio, volume filling factor, field orientation, and/or degree of alignment (e.g., Beck and Krause 2005).  
The emissivity $\ep$ is related to $L$ and $\Tsyn$ through equation (\ref{Tsynbyep}),
\be
\(\ep\/{\rm erg\ cm^{-3}s^{-1}}\) \sim 
6.796\times 10^{-12}\(\Tsyn \/{\rm K}\) \(\nu \/{\rm GHz}\)^3 \(\L\/{\rm pc}\)^{-1}.
\label{ep}
\ee 
The emissivity depends on the filling factor through $\ep \propto L^{-1}\propto \eta^{-1/3}$. So, we here introduce an $\eta$-corrected magnetic strength,
\be
\BQ\sim \eta^{2/21}\avBtot/Q.
\ee
For $\eta\sim 0.035$ as measured in the previous section, we obtain $\BQ \sim 0.73 \avBtot/Q$.    
 
Figure \ref{mapBtot} shows an all-sky map of the calculated total magnetic intensity $\avBtot/Q$. Except for discrete radio sources including radio spurs and GC, the map shows a smooth local magnetic intensity within $\sim 200$ pc. 
 Local magnetic strengths were calculated in intermediate latitude regions at $+30\deg \le b \le +70\deg$ and $-70\deg \le b \le  -30\deg$ to obtain $\Btot/Q=10.7\pm 3.1\ \muG$ and $10.3 \pm 2.7 \muG$, respectively. Combining the two regions, we obtain $\Btot/Q=10.5\pm 3.0 \muG$. 
By correcting for the volume filling factor ($\eta^{2/21}=0.73$), we obtain a representative local field strength of $\BQ=7.6\pm 2.1 \muG$ for $Q=1$.  



Given $\avBtot$ and $\avBpara$ maps, the perpendicular component of the magnetic field is easily calculated by $\avBperp\sim \sqrt{\avBtot^2-\avBpara^2}$. However, the accuracy of the above estimated $\avBtot$ would be too poor to obtain a meaningful map of the perpendicular component.

\section{Discussion}

\subsection{Summary}

The latitudinal plots in figure \ref{All-Lat} indicate that the four observables, $RM$, $\Tsyn$, $\Tff$, and $\NHI$, are tightly correlated with each other through their cosec $|b|$ variations. This indicates that the distributions of the sources and their physical parameters are also tightly correlated with each other. Based on this fact, the sources of these emissions and Faraday rotation are assumed to be distributed in a single local disk in the Galaxy.

On this assumption, some useful relations were derived for calculating the local ISM quantities such as magnetic strength $\avBtot$, and LOS component of magnetic field $\avBpara$, thermal electron density $\avne$, \textsc{HI} density $\avnHI$, and LOS depth $L$, or the scale thickness $\zhalf$ and $\zdisk$ with the volume filling factor $\eta$. It was emphasized that determination of $L$ plays an essential role in the present hybrid method to calculate the physical quantities, while only $\avBpara$ can be directly calculated from $RM$ and $\NHI$ without being affected by $L$.

Applying the method to archival radio data, all-sky maps of $\avBpara$ and $\avBtot$ were obtained, which revealed a detailed magnetic structure in the local interstellar space within the Galactic disk near the Sun. The $\avBpara$ map showed that the magnetic direction varies sinusoidally along a giant arch-shaped belt on the sky, changing its LOS direction from north to south and vise versa every two galactic quadrants. Maximum parallel component of $\sim \pm 5-6 \muG$ was observed on the belt at intermediate latitudes. The $\avBtot$ map showed that the total magnetic strength is smoothly distributed on the sky, and the averaged value was obtained to be $\avBtot/Q \sim 10.5 \muG$ in the intermediate latitude region. Assuming an equipartition factor of $Q=1$, we obtained an $\eta$-corrected field strength of $\BQ\sim 7.6\muG$ for the measured volume-filling factor of $\eta\sim 0.035$.

\subsection{Dependence on the thermal electron fraction}

The proportionality of the densities of thermal electrons and \textsc{HI} gas is confirmed through the tight correlation between $\NHI$ and $\Tff\ (EM)$ by figures \ref{All-Lat} and \ref{ISM-ISM}. On this basis, we assumed a constant thermal electron fraction of $\xe\sim 0.1$ close to the current measurement on the order of $\sim 0.08$ (Foster et al. 2013). However, $\xe$ affects the result through equation (\ref{L}), where $L \p \xe^2$ and it propagates to the other quantities as $\avBpara \p \xe^{-1}$, $\avBtot \p \xe^{-4/7}$, $\avne \p \xe^{-1}$, and $\avnHI \p \xe^{-2}$. Namely, the ISM quantities are generally proportional inversely to $\xe$, with strongest effect on $\avnHI$ and weakest on $\avBtot$.

\subsection{Uncertainty from energy equipartition}

The most uncertain point in the present analysis is the estimation of magnetic strength from the energy equipartition of cosmic-ray electrons and magnetic field. Since the equipartition factor $Q$ is still open to discussion, the obtained total magnetic intensities should be taken only as a reference to see the relative distribution of the strength on the sky. 

ALso, the single disk assumption for synchrotron and thermal components may break at high latitudes. As in figure \ref{ISM-ISM}, the plots of $\Tsyn$ against $\NHI$ and $\Tff$ bend at high latitudes, showing an order of magnitude excess at high latitudes over smooth extension from the disk component. The synchrotron excess over that expected from frozen-in assumption is about $\delta \Tsyn \sim 0.1$ K.

\begin{figure} 
\begin{center}  
\includegraphics[width=6cm]{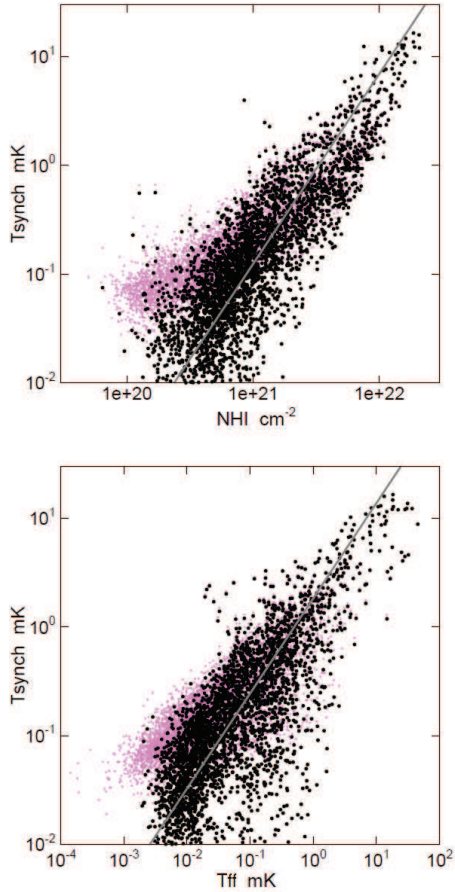}      
\end{center} 
\caption{Correlation of  $\Tsyn-0.1$ mK with $\NHI$ and $\Tff$, which are approximately represented by a power law of index 7/4 and 7/8, respectively. Vertical scaling of the lines are arbitrary. Original data are shown in violet. }
\label{syn-ism-bgrm}
\end{figure}

In figure \ref{syn-ism-bgrm} we plot $\Tsyn-\delta \Tsyn$ mimicking {\it halo-subtracted} synchrotron emission, which is well fitted by a power law expected from low and intermediate latitude regions. This fact suggests that the energy-equipartition holds inside the disk, whereas a non-thermal halo at $\sim 0.1$ mK level at 23 GHz is extending outside the gas disk. 
  
\subsection{Effect of inhomogeneity} 

From the tight cosec $|b|$ relation of the used radio observables, for which no extinction problem exists, we assumed a uniform layered disk of ISM. In more realistic conditions, however, the layer may be more or less not uniform, and the assumption made in equations (\ref{f}-\ref{fg}) may not hold, or must be modified.  

However, it is emphasized that 'clumpy' inhomogeneity does not affect the 'effective' LOS depth $L$ by definition, because $L$ already includes the volume filling factor. Hence, the determined values of the ISM, which are averages of values 'inside' the clumps (or within $L$), are not affected by the inhomogeneity. 

The assumption made for equations \ref{f} to \ref{fg} will not hold exactly in a disk with globally varying density with the height. For example, if the functions $f$ and $g$ are represented by a Gaussian function of the height from the galactic plane, we have $\la f \ra \sim  1.1 \la f^2 \ra ^{1/2}$ and $\la fg \ra \sim 1.2 \la f \ra \la g \ra$. For a cosh$^{-2}$ function, as for self-gravitating disk, the factors are 1.2 and 1.4, respectively. These factors propagate onto the results, yielding uncertainty by a factor of $\sim 1.1-1.2$ for quantities having linear dependence on the distance, and by $1.2-1.4$ to those with non-linear dependence such as $RM$ and $EM$. However, the finally determined $B$ and $\ne$ or $\nHI$ are more linearly dependent on the distance, and hence their uncertainties may be about a factor of $\sim 1.1-1.2$ at most.

\subsection{Local bubble}

\def\rbub{r_{\rm bub}}
\def\tan{{\rm tan}}

A large-scale irregularity of the ISM has been reported as a local bubble ({Bochkarev} {1992}; Lallement et al. 2003; Liu et al. 2017; Alves et al. 2018), which makes a cavity around the Sun of radius $\rbub \sim 100-200$ pc widely open to the galactic halo. Then, a difficulty is encountered to explain the tight cosec $|b|$ relation in figure \ref{All-Lat}. In order for the cosec relation to hold up to $b\sim \pm 80\deg$ at least, the scale height of the galactic disk must be greater than $\sim \rbub \tan \ 80\deg \sim $ 600 pc to 1.2 kpc, which is obviously not the case. If the disk scale height is $\sim 200$ pc as measured in HI, the open cavity should result in huge empty sky in radio around the galactic poles, which also appears not the case.

\begin{figure} 
\begin{center}   
\includegraphics[width=7cm]{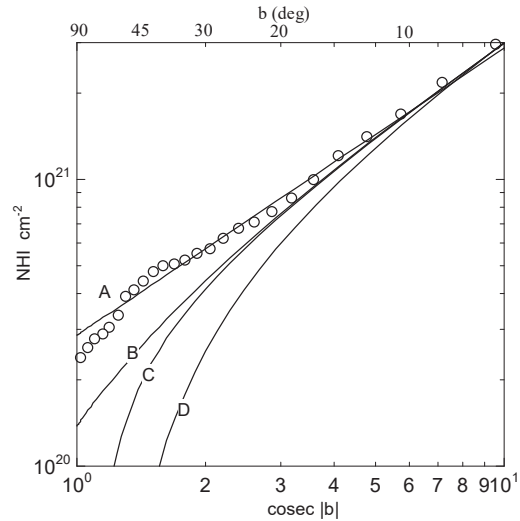}       
\end{center} 
\caption{  Cosec $|b|$ relation of the \textsc{HI} disk from figure \ref{All-Lat} (averaged values every $2\deg$ by circles) compared with (A) a model of Gaussian disk of scale height 200 pc without bubble; (B) same disk with a spherical bubble of radius 100 pc; (C) and (D) disk with vertical cylindrical cavity of radius 100 and 200 pc, respectively. Values are normalized to the observation at $b\sim 5\deg$. }
\label{bubble}
\end{figure}

In figure \ref{bubble} we compare the cosec $|b|$ relation observed for $\NHI$ with cavity models mimicking the local bubble. Line A indicates a Gaussian disk of scale height 200 pc without bubble; B represents a case with a spherical bubble of radius 100 pc in the Gaussian disk of scale height 200 pc, and C and D for cylindrical cavity of radius 100 and 200 pc, respectively. Model C may be compared with the result by Lallement et al. (2003), which appears significantly displaced from the cosec $|b|$ relation. Such is found not only in HI, but also in thermal and synchrotron emissions, and Faraday RM (figure \ref{All-Lat}). Therefore, the relation between the local bubble and the cosec $|b|$ disk in radio remains as a question. 
  
\subsection{Other observables}

Molecular gas has not been taken into account in this study, because the nearest molecular clouds within LOS depths concerned in this paper are rather few (Knude and Hog 1998). Comparison with a local bubble surrounded by dusty clouds (e.g., Lallement et al. 2003), besides the cosec $|b|$ problem, would be an interesting subject, although the present analysis gives only averaged values along the LOS within $L$, and hence cannot be directly compared with the 3D study. 

Polarization data in radio and infrared observations were not used, although they are obviously useful to improve the present hybrid analysis. Inclusion of these observables is beyond the scope of this paper, for which more sophisticated analyses would be required.
 
\subsection{Local magnetic topology}

Despite of the various uncertainties as above, we emphasize that the projected topology of $\Bpara$ mapped in figure \ref{mapBpara} is rather certain. Although the {\Large $\propto$}-shaped variation of $RM$ might sound a bit strange, we could speculate possible topology of the magnetic lines of force in the local space. 

\begin{figure} 
\begin{center}  
\includegraphics[width=7cm]{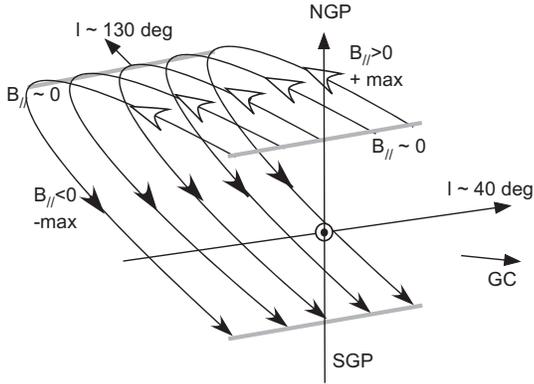}          
\end{center} 
\caption{ Possible topology of local magnetic lines of force to explain $\Bpara$ map in figure \ref{mapBpara}.  }
\label{magtopo}
\end{figure}

The field direction reverses about the galactic plane from north to south in a wide range from $l\sim  30\deg$ to $\sim 210\deg$. The $\Bpara$ value attains its maximum and minimum at both intermediate latitudes around $(l,b)\sim (130\deg,\pm 30\sim 50\deg)$. Such $RM$ behavior on the sky could be explained by a reversed topology of local field as illustrated in figure \ref{magtopo}.

\section*{Acknowledgments} 
We thank the authors of the LAB \textsc{HI} survey (Dr. Kalberla et al. ), all-sky rotation measure map (Dr. Taylor et al. ), and the WMAP 7 years maps (Dr. Gold et al.) for the archival data. The data analyses were performed on a computer system at the Astronomical Data Center of the National Astronomical Observatories of Japan.

\begin{appendix}
\section{Equipartition}

\subsection{$B$, $E$, and $N(E)$ by $\ep$ and $\nu$}

\def\me{m_{\rm e}}   
\def\nuG{\nu \/{\rm GHz}} 
\def\eperg{\(\epsilon\/{\rm erg\ cm^{-3}s^{-1}}\)}
\def\Inu{I_{\rm \nu}}

The equipartition between magnetic and cosmic-ray pressure relates the magnetic strength $B$, representative energy $E$ and density $N(E)$ of cosmic-ray electrons responsible for synchrotron emission at observing frequency $\nu$ and volume emissivity $\epsilon$ (Burbidge 1956; Moffet 1975; Sofue et al. 1986 for review). We here write down the basic relations among $B$, $N(E)$ and $E$, which are used to calculate the 'reference value' of $B$ for $Q=1$ in equation \ref{B}.
\be
{B^2\/8\pi} \sim EN(E),
\label{Beq}
\ee 
\be
\nu\sim  {e\/4 \pi \me^3 c^5} B E^2,
\label{nu}
\ee
and
\be
\epsilon\sim -{dE\/dt}N(E) \sim {2 e^4 \/3 \me^4 c^7} B^2E^2 N(E)
\label{emi}
\ee
(e.g., Landau and Lifshitz 1971).
These equations can be solved for $B$ in terms of $\nu$ and $\epsilon$ as
\be
\({B\/\muG}\) \sim 3.03 \times 10^8 \(\nuG \)^{-1/7}\eperg^{2/7}.
\label{AppB}
\ee
Moffet (1973) gave a coefficient $3.25\times 10^8$ for a radio spectral index of $\alpha=-0.75$.  

As a byproduct, we obtain
\be
\({N(E)\/{\rm cm}^{-3}}\)\sim 2.91\times 10^9 \(\nuG\)^{-6/7} \eperg^{5/7},
\ee
and
\be
\({E\/{\rm erg}}\)\sim 1.26\times 10^{-6}\(\nuG\)^{4/7}\eperg^{-1/7}.
\ee 

\subsection{$\Tsyn$   by $\NHI$}

Using equations (\ref{Sigma}), (\ref{Beq}) and (\ref{nu}), $\Tsyn$ is expressed in terms of $L$ and $B$ as
\be
\Tsyn \propto LB^{~9/2}.
\label{TsynB}
\ee
Assuming that the \textsc{HI} gas has a constant velocity dispersion ($\sigma\sim$ constant) and is in pressure (energy density) balance with magnetic field as
\be
{B^2\/8\pi} \sim {1\/2}\mH \nHI \sigma^2 \propto \nHI,
\ee
we have
\be
B\propto \nHI^{1/2}\sim (\NHI/L)^{1/2}.
\ee
Inserting this to equation (\ref{TsynB}), 
\be
\Tsyn \propto \nHI^{1/4}\NHI^{7/4}.
\ee
While column density $\NHI$ is highly variable with $b$ and $L$, the volume density $\nHI$ is not, and appears by a weak power of index 1/4. So, we may approximate $\Tsyn$ by
\be
\Tsyn \propto \NHI^{7/4}.
\ee

\end{appendix} 

\end{document}